# RIPL: An Efficient Image Processing DSL for FPGAs

Robert Stewart & Greg Michaelson
Mathematical & Computer Sciences
Heriot-Watt University
Edinburgh, UK
{R.Stewart,G.Michaelson}@hw.ac.uk

Deepayan Bhowmik & Andrew Wallace
Engineering & Physical Sciences
Heriot-Watt University
Edinburgh, UK
{D.Bhowmik,A.M.Wallace}@hw.ac.uk

*Abstract*—Field programmable gate arrays (FPGAs) can accelerate image processing by exploiting fine-grained parallelism opportunities in image operations. FPGA language designs are often subsets or extensions of existing languages, though these typically lack suitable hardware computation models so compiling them to FPGAs leads to inefficient designs. Moreover, these languages lack image processing domain specificity. Our solution is RIPL[1], an image processing domain specific language (DSL) for FPGAs. It has algorithmic skeletons to express image processing, and these are exploited to generate deep pipelines of highly concurrent and memory-efficient image processing components.

## I. Introduction

FPGAs can be configured directly with hardware description languages, though these require hardware expertise and come with the cost of long debugging stages to remove design errors. Alternatives include high level synthesis tools to compile existing imperative languages, and dataflow languages that abstract the highly concurrent nature of FPGA hardware. However, the absence of suitable hardware to support the imperative model make compiling them to FPGAs very inefficient, and dataflow languages burden programmers with wiring together computations explicitly.

RIPL abstracts dynamic dataflow process networks (DPNs) by hiding actors and wires, and inherits DPN hardware abstractions of clocks, signals, registers and memory. The RIPL programmer uses a collection of built in image processing skeletons, and the compiler automatically extracts parallelism from the program, to generate deeply pipelined and memory efficient FPGA designs.

## II. Design

### A. Requirements & Constraints

Higher order computer vision algorithms are composed of lower level image operations. Prototypical image processing operations can be classified in terms of the locality of their data access requirements: pixel to pixel functions on *points*, neighbourhood pixels to pixel functions on *regions*, and *global* operations on entire images.

The memory constraints of FPGAs mean that many CPU & GPU methods for parallel image processing cannot be adopted for FPGA image processing implementations. Software techniques often store arrays whose sizes matches complete images, and apply data-parallel kernels in a vectorised single instruction multiple data (SIMD) or coarse grained single instruction multiple threads (SIMT) fashion. These image processing models are prohibitive for FPGA implementations, because on-chip memory is a very scarce resource, and the global shared memory model is not suitable for the inherently fine grained concurrent nature of FPGAs. Modern CPUs have access to around 2MB cache and 64GB of RAM, which are treated as a large shared memory block. In contrast for example, a modern Virtex 7 FPGA has a total of just 8.5MB of available on-chip block RAM (BRAM) memory.

### B. RIPL Overview

RIPL is a functional language with single assignment semantics. It comprises domain specific image processing types, functions and algorithmic skeletons [1]. RIPL's algorithmic skeletons are reusable parameterised descriptions of task-specific image processing architectures and are exploited to generate pipelines of image operations. The skeletons process pixel vectors in rows, columns and regions with computation kernels, which are lightweight functions that traverse over images.

An illustration of pipelined skeleton composition is in Figure 1. The RIPL skeletons API is shown in Figure 2, using standard notation for function type signatures, *e.g.* `mapRow` takes as arguments: an $M \times N$ image, a function from a vector of $A$ pixels to a vector of $A$ pixels, and returns an $M \times N$ image. The `map` skeletons are element or column/row wise mappings from pixels to pixels. The `zipWith` skeleton takes two images and merges them into a single stream with some user defined merging function. The `combine` skeleton takes entire rows or columns from two images and merges them into a single stream with built-in RIPL operator, such as `append`. The `convolve` skeleton is parameterised by a window dimension and computes pixel values from a neighbourhood of pixels. The `fold` skeleton is parameterised by an initial value and applies global operations over an image and returns a scalar value or a vector.

RIPL uses index types to impose the constraint that all skeletons operate on images with bounded shapes known at compile time. For example, an inferred indexed data type $Im_{(50,40)}$ is an image of width 50 and height 40, and $[P]_8$ is a vector of 8 pixels. This allows the RIPL compiler to generate actors with static arrays from these indexed data structures, enabling HDL synthesis tools to make optimal memory implementation choices about static structures, *i.e.* look-up tables (LUTs) for small arrays or with combined BRAM blocks for larger arrays.

---

[1]Rathlin Image Processing Language

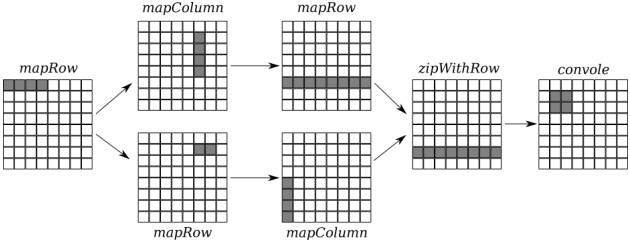

Fig. 1. Pipelining skeleton compositions

$$mapRow : Im_{(M,N)} \to ([P]_A \to [P]_A) \to Im_{(M,N)}$$
$$mapCol : Im_{(M,N)} \to ([P]_A \to [P]_A) \to Im_{(M,N)}$$
$$concatMapRow : Im_{(M,N)} \to ([P]_A \to [P]_B) \to Im_{(B/A*M,N)}$$
$$concatMapCol : Im_{(M,N)} \to ([P]_A \to [P]_B) \to Im_{(M,B/A*N)}$$
$$zipWithRow : Im_{(M,N)} \to Im_{(M,N)} \to (P \to P \to P) \to Im_{(M,N)}$$
$$zipWithCol : Im_{(M,N)} \to Im_{(M,N)} \to (P \to P \to P) \to Im_{(M,N)}$$
$$combineRow : Im_{(M,N)} \to Im_{(M,N)}$$
$$\to ([P]_A \to [P]_A \to [P]_B) \to Im_{(B/A*M,N)}$$
$$combineCol : Im_{(M,N)} \to Im_{(M,N)}$$
$$\to ([P]_A \to [P]_A \to [P]_B) \to Im_{(M,B/A*N)}$$
$$convolve : Im_{(M,N)} \to (a,b) : (Int, Int)$$
$$\to ([P]_{a*b} \to P) \to Im_{(M,N)}$$
$$foldVector : Im_{(M,N)} \to Int \to s : Int$$
$$\to (P \to [Int] \to [Int]) \to [Int]_s$$
$$foldScalar : Im_{(M,N)} \to Int \to (P \to Int \to Int) \to Int$$

Fig. 2. RIPL skeletons API

## III. Implementation

### A. RIPL to Dataflow

The RIPL compiler uses the dynamic dataflow process network (DPN) model as an intermediate representation between the DSL and FPGA implementations. To address FPGA memory limitations, the compiler eliminates intermediate image arrays by feeding rows and columns through concurrent phases of a pipeline. The kernel at each phase is provided only the pixel values they need to execute. Costly intermediate arrays are therefore avoided for local and regional data access patterns with RIPL skeletons.

Thanks to RIPL's single assignment semantics, implicit data dependencies in skeleton compositions are exploited to generate deeply pipelined graphs from RIPL programs. The vertices (actors) represent image operations and the edges (wires) represent dataflow between composed operations. Transposition actors are added whenever a row wise skeleton is composed with a column wise skeleton, and vice versa. Informally, the mapping from skeletons to graphs is as follows. A skeleton instance maps to one actor. The arity of a skeleton maps to the number of input ports the corresponding actor has. The number of output ports of an actor is dictated by the number of other skeletons that use the output image of the skeleton. Implicit dataflow in the composition of skeletons is lifted to explicit wires between actors. The user defined function to a skeleton becomes a fireable rule inside the actor. The graph is mapped onto FPGAs to exploit two kinds of pipelined parallelism: 1) to feed the rows/columns of an image through different pipeline stages, and 2) to feed multiple video frames into the FPGA fabric concurrently.

### B. Dataflow to FPGAs

The generated dataflow graph is expressed with the CAL actor language [2]. An existing CAL to Verilog compiler [3] is used to add an interface protocol for actor interconnects and an explicit clock to all actor components, then lowers the graph to an FPGA abstraction adding signals, registers, FIFOs and shared memories. Generic memories are used for arithmetic operations, registers and actor interconnects, allowing HDL synthesisers to choose from LUT or BRAM instantiations, depending on holistic memory requirements and on the FIFO depths needed to support implicit dataflow dependencies in RIPL programs.

## IV. Discussion & Conclusion

In this abstract we present RIPL, a high level image processing DSL for FPGAs. It has high level image processing skeletons familiar to software programmers, which are exploited to generate deep pipelines of memory-efficient image processing operations. RIPLs underlying dynamic dataflow model supports different image data access patterns using skeletons. The aim of RIPL is to maximise clock frequency to increase throughput, and to minimise BRAM use to fit complex algorithms onto FPGAs. RIPL has been used to implement image watermarking and multi-dimensional subband decomposition algorithms. We believe that RIPLs underlying dynamic DPN semantics provides greater levels of expressivity compared to other image processing FPGA languages. Ongoing work includes evaluating the expressivity of RIPL with a comprehensive collection of case studies. We plan on integrating RIPL with a performance guided dataflow transformations framework we are developing [4].

### Acknowledgements

We acknowledge the support of the Engineering and Physical Research Council, grant references EP/K009931/1 (Programmable embedded platforms for remote and compute intensive image processing applications).